\documentclass[manuscript]{emulateapj}
\usepackage{amsmath}

\shorttitle{Precise Measurement of Reionization Optical Depth from Global 21-cm  with Realistic Heating}
\shortauthors{Fialkov \& Loeb}

\begin{document}


\title{Precise Measurement of the Reionization Optical Depth from The Global 21-cm Signal Accounting for Cosmic Heating}


\author{Anastasia Fialkov}
\affil{Department of Astronomy, Harvard University,
60 Garden Street, $MS-51$, Cambridge, MA, 02138 U.S.A.}
\email{anastasia.fialkov@cfa.harvard.edu}
\and
\author{Abraham Loeb}
\affil{Department of Astronomy, Harvard University,
60 Garden Street, $MS-51$, Cambridge, MA, 02138 U.S.A.}
\email{aloeb@cfa.harvard.edu}


%\altaffiltext{1}{Visiting Astronomer, Cerro Tololo Inter-American Observatory.
%CTIO is operated by AURA, Inc.\ under contract to the National Science
%\altaffiltext{2}{Society of Fellows, Harvard University.}
%\altaffiltext{3}{present address: Center for Astrophysics,
%    60 Garden Street, Cambridge, MA 02138}
%\altaffiltext{4}{Visiting Programmer, Space Telescope Science Institute}
%\altaffiltext{5}{Patron, Alonso's Bar and Grill}


\begin{abstract}

As a result of our limited data on reionization, the total optical depth for electron scattering,  $\tau$, limits precision measurements of cosmological parameters from the  Cosmic Microwave Background (CMB).  It was recently shown that the predicted 21-cm signal of neutral hydrogen contains enough information to reconstruct $\tau$ with sub-percent accuracy, assuming that the  neutral gas was much hotter than the CMB throughout the entire epoch of reionization. Here we relax this assumption and use the global 21-cm signal alone to extract $\tau$ for realistic X-ray heating scenarios. We test our model-independent approach using mock data for a wide range of ionization and heating histories and show that an accurate measurement of the reionization optical depth  at a sub-percent level is possible in most of the considered scenarios even when heating is not saturated during the  epoch of  reionization, assuming that the foregrounds are mitigated. However, we find that in cases where heating sources had hard X-ray spectra  and their luminosity was close to or lower than what is predicted based on low-redshift observations, the global 21-cm signal alone is not a good tracer of the reionization history. 

\end{abstract}



\section{Introduction}

The reionization of the intergalactic medium (IGM), between redshifts $z\sim 13$  and $6$ \citep{Zahn:2012,George:2015,Ade:2015,Becker:2015}, is one of the least studied  epochs in the history of the Universe and is a  research frontier in present-day cosmology \citep{Loebbook}. During this process, the neutral intergalactic gas was likely ionized by ultra-violet (UV) photons emitted by stars. In addition, sources of X-ray photons, such as X-ray binaries, mini-quasars and hot gas in galaxies, also had an effect on the epoch of reionization (EoR) by pre-heating and mildly ionizing the gas far from the sources \citep{Oh:2001, Mesinger:2013, Fialkov:2015}.  

The EoR also affects the  Cosmic Microwave Background (CMB) through its scattering off  free electrons. This scattering degrades the accuracy with which cosmological parameters can be extracted from the CMB data. In particular, measurements of the   amplitude of primordial  fluctuations, $A_s$, is degenerate with the total optical depth, $\tau$, since the total amplitude is estimated from the temperature power spectrum of the CMB as $A_s e^{−2\tau}$.  Because the precision with which $\tau$ can be measured using the CMB is very poor (e.g., the 68\% confidence level in $\tau$ corresponds to a relative error of $\sim 24\%$ when measured from the temperature power spectrum), the errors in $A_s$ are high.   As a result, $\tau$ is  sometimes referred to as a nuisance parameter for CMB cosmology. Luckily, alternative probes of reionization can provide independent constraints on $\tau$ and remove the related uncertainty.

One of the most promising tools to constrain reionization is the predicted 21-cm signal of neutral hydrogen, HI, e.g., see \citet{Furlanetto:2006b} and \citet{Pritchard:2012}. The brightness temperature of this signal, $\delta T_b(z)$,  depends on the fractional amount of hydrogen atoms in the IGM which are neutral, $x_{HI}$, and, thus, is expected to provide exclusive information on the reionization history of the Universe. Recently, \citet{Liu:2015}  advocated  that the sky-averaged (global) 21-cm signal, $\overline{\delta T_b(z)}$, alone has enough information to fully reconstruct the reionization history and measure the optical depth to reionization with great precision. To alleviate the computational costs, the authors assumed that  the 21-cm signal tracks the ionization history, which is true only when X-ray  sources  heat up the cosmic gas to a temperature above the CMB well before the beginning of reionization. In this case, the dependence of $\delta T_b(z)$ on the gas temperature is saturated (the so-called  saturated heating regime),  $\delta T_b(z)$    is proportional to $x_{HI}$, and the reionization history can be fully extracted from the global 21-cm signal measurements despite the presence of strong foregrounds. In particular,  \citet{Liu:2015} showed that assuming saturated heating, the 21-cm signal allows to determine $\tau$ with much higher accuracy than it is  possible from the CMB measurements.

However, the assumption that heating is saturated all the way through reionization is debated \citep{Fialkov:2014N}, and the nature and efficiency of early X-ray sources could have a significant impact on the intensity of the redshifted 21-cm signal even at the end of the EoR \citep{Pritchard:2012, Mesinger:2013, Fialkov:2014N, Pacucci:2014}. The nature of the first X-ray sources is still unknown and possible candidates   include X-ray binaries \citep{Mirabel:2011,Fragos:2013,Fialkov:2014N} and mini-quasars \citep{Madau:2004, Fialkov:2015} which emit hard  X-rays with  spectral energy distribution (SED) peaking at few keV, soft X-ray sources such as hot gas in galaxies which can be well  described by a power-law spectral shape \citep{Furlanetto:2006}, as well as more exotic candidates such as annihilating dark matter \citep{Cirelli:2009}. The efficiency of high-redshift sources, $f_X$, defined through the relation between their luminosity and the star formation rate  is another unknown (\citet{Fialkov:2015} and references therein), calibrated so that the value of $f_X =1$ corresponds to the luminosity of observed low-redshift sources boosted by a metallicity-dependent factor  \citep{Fragos:2013}.

Here we consider a completely model-independent method to reconstruct $\tau$ from the global 21-cm signal measurements after relaxing the saturated heating assumption and examining realistic  X-ray sources with hard and soft spectra varying their efficiency.  Our results are timely since many of the experiments such as  the Experiment to Detect the Global Epoch of Reionization Signature (EDGES, \citet{EDGES}), Large-Aperture Experiment to Detect the Dark Ages (LEDA, \citet{LEDA}, \citet{Bernardi:2015}), Dark Ages Radio Explorer (DARE, \citet{DARE}), and New extension in Nancay upgrading LOFAR (NenuFAR, \citet{NenuFAR}) are on their way to detect  this signal for the first time while next generation telescopes, such as the Hydrogen Epoch of Reionization Array (HERA\footnote{http://reionization.org/}) and the Square Kilometer Array (SKA, \citet{Koopmans:2015}), are expected to extensively explore the EoR.

In Section \ref{Sec:sim} we set up the stage describing simulation methods and model parameters. In Section \ref{Sec:rec} we explore to which extent the global 21-cm signal tracks the neutral fraction in each case and propose a model-independent way to reconstruct the heating and ionization history from the global 21-cm signal. In Section \ref{Sec:OD} we calculate the optical depth from the reconstructed reionization history   and discuss the accuracy with which it can be detected by global 21-cm experiments. Finally, we conclude in Section \ref{Sec:conc}. Throughout this paper we assume the standard Planck satellite cosmology \citep{Ade:2015}.

\section{The Mock Universe} 
\label{Sec:sim}
We simulate the mock global 21-cm signal from the redshift range $z = 6-40$ using a hybrid simulation, first introduced by \citet{Visbal:2012} and described in more detail by \citet{Fialkov:2014N}. This simulation allows to estimate the non-local impact of  X-ray, Ly-$\alpha$ and UV sources on the redshifted  21-cm signal of neutral hydrogen as well as on the   ionization history of the IGM, and includes the effect of supersonic flows between dark matter and gas, $v_{bc}$ \citep{Tseliakhovich:2010},  which has an impact on high-redshift star formation  in $10^5-10^7$ M$_\odot$ halos \citep{Stacy:2011, Maio:2011} and, consequently, on the 21-cm signal \citep{Dalal:2010, Tseliakhovich:2011, Fialkov:2012, McQuinn:2012, Fialkov:2014R}. In addition, we account for the photoheating feedback \citep{Cohen:2015} which happens when the intergalactic gas heats up and  stops  accreting into  halos below $\sim  10^8 - 10^9$ M$_\odot$, thus  suppressing star formation in low-mass halos. 

In our simulation, ionization by UV photons is computed following the  excursion-set formalism, by comparing the time-integrated number of ionizing photons to the number of neutral atoms in each region  \citep{Furlanetto:2004}. Specifically, a simulation cell is  ionized if $\zeta_{UV}f_{coll} \geq (1-x_e)$, where $\zeta_{UV}$ is the ionization efficiency normalized to yield $\tau$,  $f_{coll} $ is the collapsed fraction, and $x_e$ is the fraction of free electrons. In addition, we account for partial ionization of the neutral IGM by X-rays, which boost the free electron fraction far from the sources and have a non-negligible effect on the topology of reionization. 

The reionization history is strongly linked to the mechanism of star formation and its timing depends on the minimal mass of halos that can form stars, M$_{\textrm{min}}$. The smaller is  M$_{\textrm{min}}$, the earlier reionization starts and the more gradual is the grows of the ionized fraction. Because star formation at high redshifts is very unconstrained and is biased by multiple feedback mechanisms \citep{Greif:2015, Bromm:2013}, we consider three different scenarios varying the low-mass cutoff of star forming halos:
\begin{itemize}
\item ``Massive halos'': Stars form in halos of M$_{\textrm{min}}\gtrsim 10^9$ M$_\odot$ (circular velocity $\geq 35.5$ km/s).
\item ``Atomic cooling'': Stars form through the atomic cooling channel in halos of M$_{\textrm{min}}\gtrsim 10^7$ M$_\odot$  (circular velocity $\geq 16.5$ km/s) with active photoheating feedback.  
\item ``Molecular cooling'': star formation happens in all halos with circular velocity $\geq 4.2$ km/s (M$_{\textrm{min}}\gtrsim 10^5$ M$_\odot$). In this case we include the photoheating feedback, account for the effect of $v_{bc}$, but exclude the effect of Lyman-Werner (LW) photons which are expected to destroy molecular hydrogen acting as negative feedback to star formation. The degree to which the LW feedback is efficient is a topic of active research \citep{Schauer:2015, Visbal:2014}; therefore, we ignore the effect of this feedback here to optimize the contribution of the molecular cooling halos and increase the diversity of ionization histories. The case of molecular cooling with LW and  $v_{bc}$ included is close to the atomic cooling scenario \citep{Fialkov:2013} which we consider separately. Although the role of molecular cooling halos in reionization is expected to be small based on  the low optical depth found by Planck satellite, their contribution  is not ruled out considering large uncertainty in  $\tau$ measurements. 
\end{itemize}
In all the above cases we  assume a star formation efficiency of $f_\star = 5\%$.  We  consider the contribution of hydrogen and first helium reionizatio to $\tau$, assuming that singly ionized helium and hydrogen are ionized to the same fraction, $x_e$ \citep{Wyithe:2003}, and  normalize our models to yield $\tau$ consistent with Planck \citep{Ade:2015}  while also requiring reionization to end by $z=6$ or earlier.  For atomic and massive halos we choose  $\tau = 0.082$  which gives the redshift of full reionization, $z_r$,  being   $z_r \sim 6.5$ and $z_r\sim  8$ respectively. This value of $\tau$  is between the Planck and WMAP measurements of the optical depth and is $1\sigma$ away from the Planck's best fit value of 0.066. In the case of molecular cooling the process of reionization is very gradual, and we need to take $\tau= 0.114$ ($3\sigma $ away from the Planck's best fit value) to have reionization end by $z_r\sim 6$. 

Finally, we consider two types of heating sources: (i) X-ray binaries with hard SED, and (ii) soft sources with power-law SED (as described by \citet{Fialkov:2014N}). In addition, we consider three different values of heating efficiency for each type of sources: $f_X = 0.3$ (low), $f_X = 1$ (standard) and $f_X = 30$ (high). The choice of low and high heating efficiencies is motivated by rather poor observational constrains on the temperature of the IGM before the end of reionization. The unresolved soft cosmic X-ray background, which amounts to $\sim 25\%$ of the flux in the $0.5-2$ keV {\it Chandra}  band \citep{Lehmer:2012}, sets an upper limit on $f_X$  when attributed to the high redshift sources \citep{Dijkstra:2012, Mesinger:2013, Fialkov:2015}. Depending on the details of star formation and for EoR ending at $z_r\sim 6$ this measurement yields an upper bound of $f_X\sim 16-36$ ($f_X \sim 45-75$) in the case of  hard (soft) X-rays; while for $z_r\sim 8.5$  the efficiencies should be   $\sim 5$ times higher \citep{Fialkov:2015}. Here we choose $f_X = 30$ as a representative value of the high heating efficiency for all the considered models.  The lower limit on $f_X$ comes from the data collected by the Precision Array for Probing the EoR (PAPER, \citet{Pober:2015}, \citet{Ali:2015}) which rules out 21-cm fluctuations of power greater than $\sim 500$ mK$^2$ at $z = 8.4$ in the  k = 0.15-0.5 h Mpc$^{-1}$ range, where h is the Hubble constant in units of 100 km s$^{-1}$ Mpc$^{-1}$. This constrain translates into  $f_X\gtrsim 0.01$ (0.001) for hard (soft) X-ray sources in the atomic cooling case. However, for such low efficiency, the gas appears to be colder than the CMB by the end of EoR, and the method which we present in this paper does not apply. Therefore we choose $f_X = 0.3$ as our low heating efficiency value.
% and   $f_X\gtrsim 0.2$ (0.03) for hard (soft) sources for massive halos only. 

For every model we output global neutral fraction, $\bar x_{HI}$ (which we refer to as the true reionziation history), average kinetic gas temperature $T_K$ and the 21-cm signal which  depends on the ionization and thermal history   in the following way,  
\begin{equation}
\delta T_b \approx\delta T_{b,0}(1+z)^{1/2}x_{HI}(1+\delta)\left(1-\frac{T_{CMB}}{T_S}\right),
\label{Eq:2}
\end{equation}
where $\delta T_{b,0}$ is a constant that depends on atomic physics and cosmological parameters,  $\delta$ is the baryon overdensity which is statistically known  from cosmology, and $T_{CMB}$ is the CMB temperature. Here we ignore the peculiar velocity term, which adds a small correction to the global 21-cm signal \citep{Bharadwaj:2004, Barkana:2005}. Finally,  $T_S$ is the spin temperature of the 21-cm transition which depends on environment. In particular, when Ly-$\alpha$ coupling is saturated, which is usually true for $z < 25$,   we can equate the spin temperature to  gas kinetic temperature, $T_S \approx T_K$ \citep{Madau:1997}; while $T_S\rightarrow 1$  when the  IGM is much hotter than the CMB  (the saturated heating case). In the latter case Eq. (\ref{Eq:2}) can be further simplified, $\delta T_b  \propto (1+z)^{1/2} x_{HI}(1+\delta)$,  and the 21-cm signal can be used as a tracer of neutral fraction weighted by the density fluctuations. 

\begin{figure*}
\includegraphics[width=3.4in]{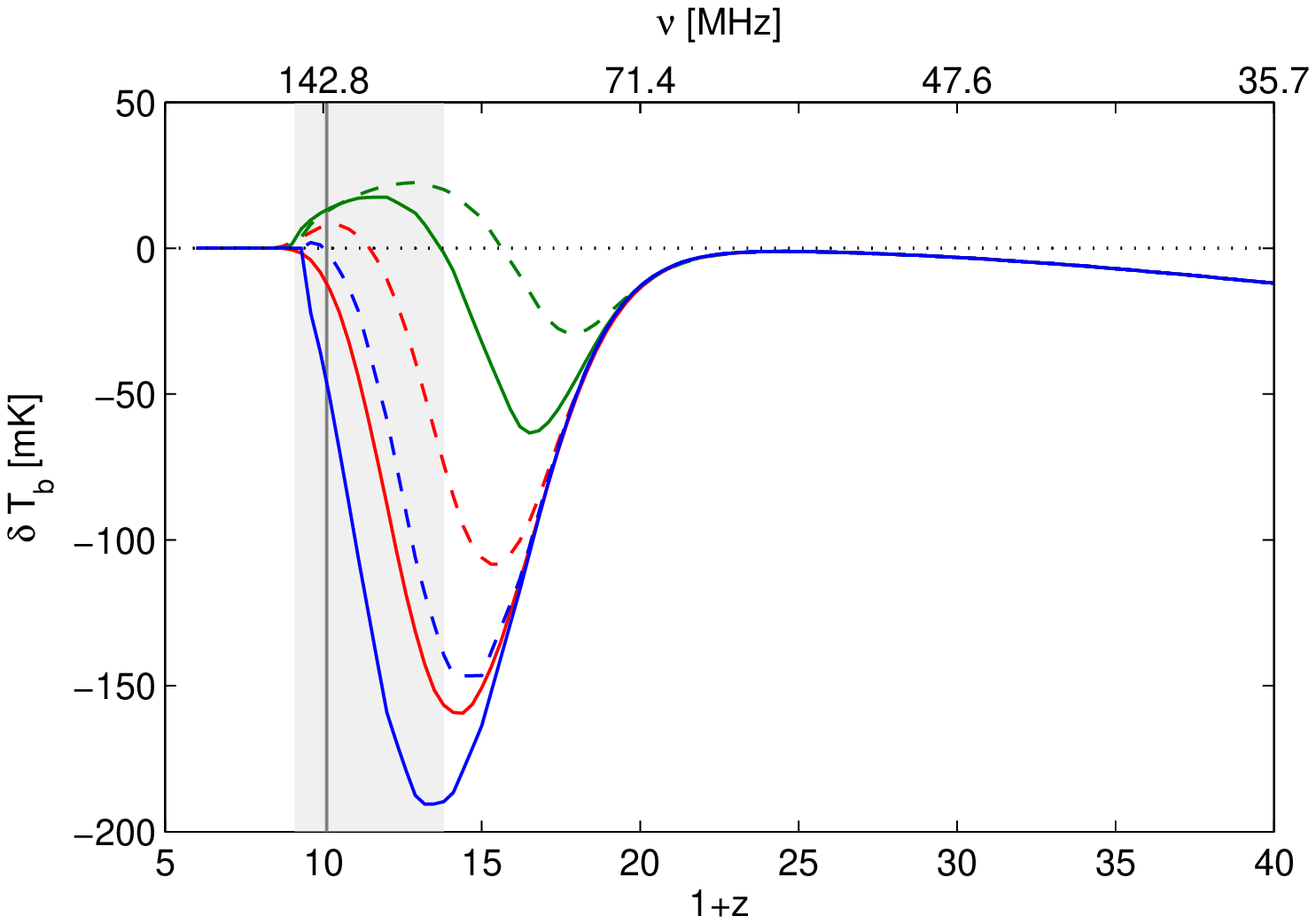}\includegraphics[width=3.4in]{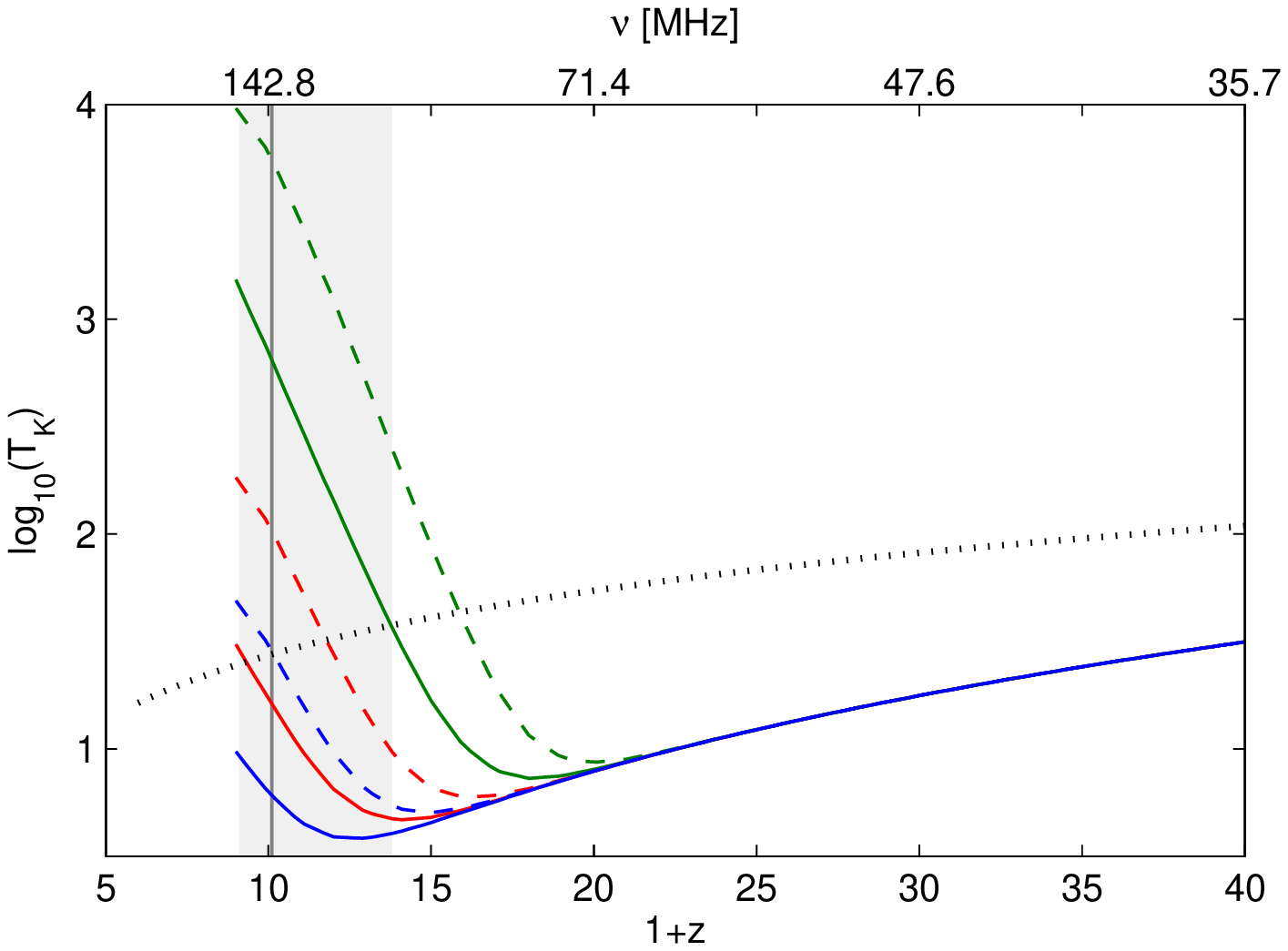}
\includegraphics[width=3.4in]{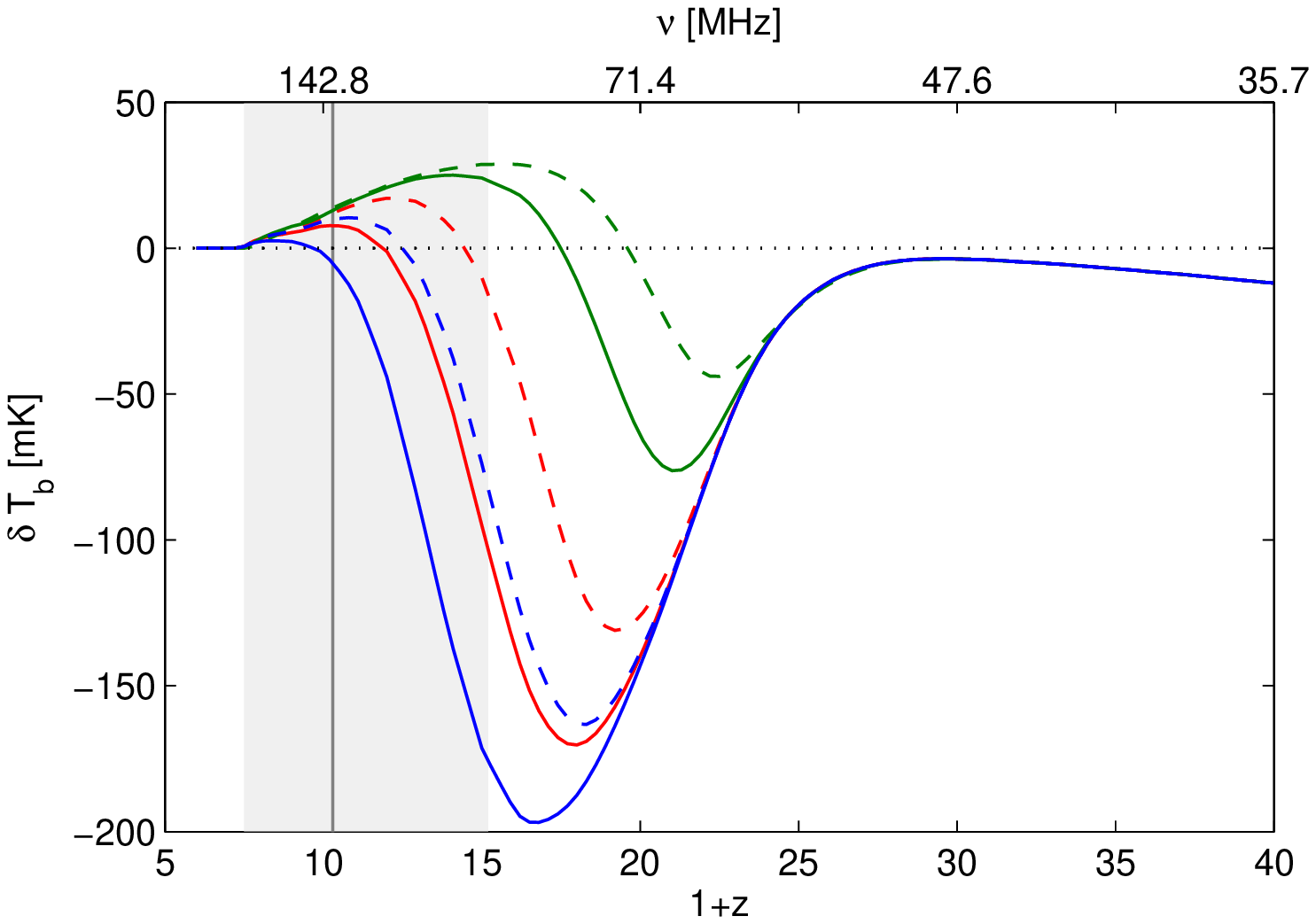}\includegraphics[width=3.4in]{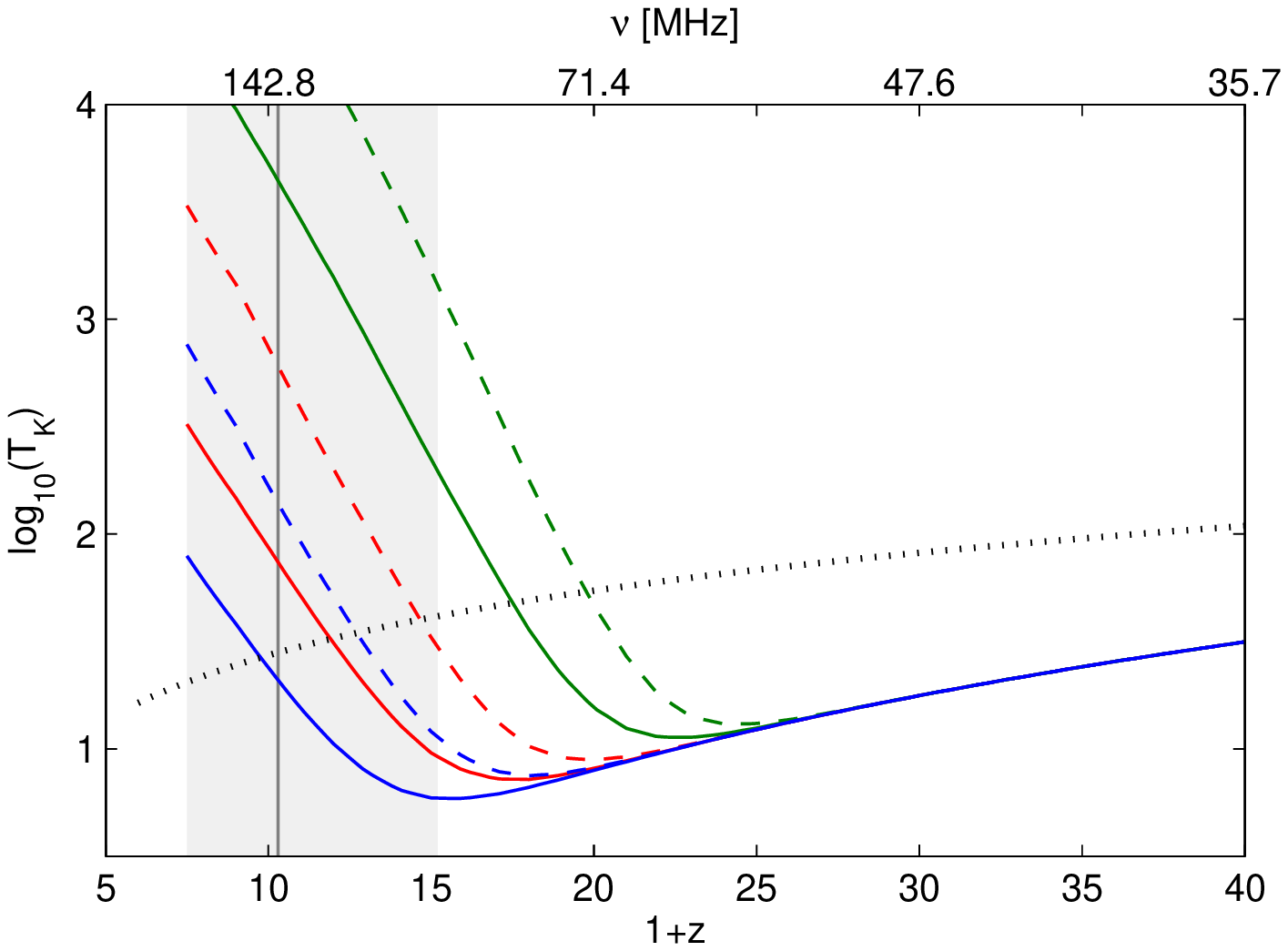}
\includegraphics[width=3.4in]{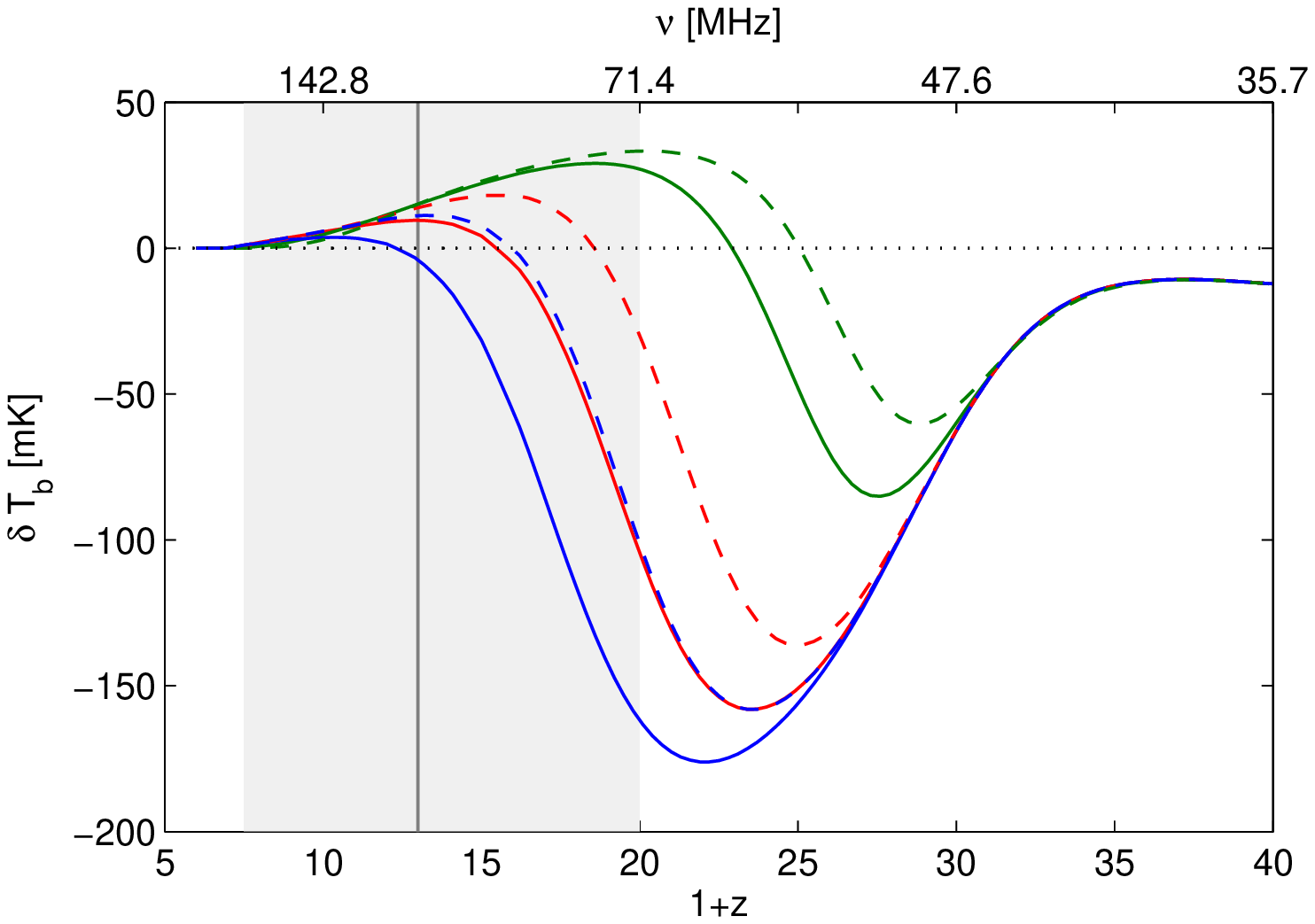}\includegraphics[width=3.4in]{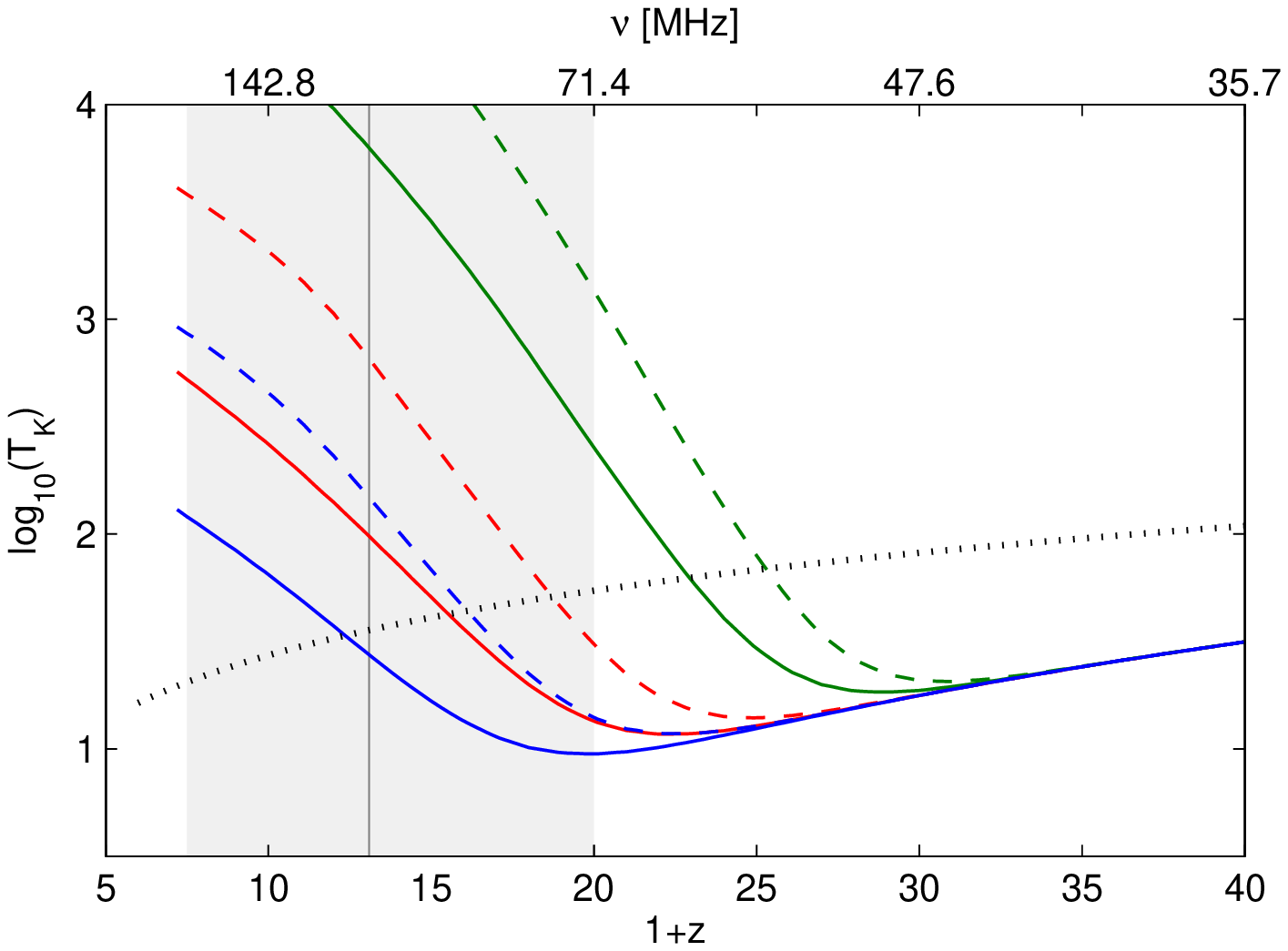}
\caption{{\bf Left:} Global 21-cm signal for all the considered models:  massive halos (top panel), atomic cooling (middle panel) and molecular cooling (bottom panel) are shown for the cases of hard SED (solid) and soft SED (dashed) for $f_X = 0.3$ (blue), $f_X = 1$ (red) and $f_X = 30$ (green).  The grey band marks the EoR from $\bar x_{HI} = 0.95$ to $\bar x_{HI} = 0.05$ and the vertical line marks the middle point of the EoR ($\bar x_{HI} = 0.5$). {\bf Right:} Kinetic gas temperature of the IGM  for the models shown on the left (same color code).  The dotted line is the temperature of the CMB. }
\label{Fig:1}
\end{figure*}

Typical global spectrum of the 21-cm signal (left column of Figure \ref{Fig:1}) features a prominent trough  at frequencies corresponding to redshifts where the IGM was colder than the CMB (the signal is seen in absorption). The minimal value of $\overline{\delta T_b}$ is reached at the beginning of heating era at redshift $z_{\textrm{min}}$ when the first population of X-ray sources turned on. At this point also the temperature of the gas, which was adiabatically cooled by cosmic expansion,  reaches its minimum  (right column of Figure \ref{Fig:1}). X-ray sources inject energy into the IGM heating it up and above the temperature of the CMB, if heating is sufficiently strong. In this case the 21-cm signal is seen in emission against the CMB at redshifts lower than $z_0$ where $T_K = T_{CMB}$. The emission signal peaks at $z_{\textrm{max}}$ and its amplitude declines at lower redshifts as reionization progresses. If heating is not  strong enough, pockets of neutral gas remain colder than the CMB throughout the EoR, marked by a grey band in each panel of Figure \ref{Fig:1}, and the 21-cm signal is seen in absorption all the way down to $z_r$.  We list $z_{\textrm{min}}$, $z_0$ and $z_{\textrm{max}}$ in Table 1 for every considered model.

\begin{deluxetable*}{llllllllllllll}
\tabletypesize{\scriptsize} \tablecaption{Summary of the results for each structure formation (column 1) and heating (column 2) model. First, we summarize the critical points of the  global 21-cm signal: the redshift at which the signal is minimal ($z_{\textrm{min}}$, column 3), vanishes ($z_0$, column 4) and is maximal ($z_{\textrm{max}}$, column 5). Next, we note the value of  $\overline x_{HI}$ at the point in time when  $\bar x_{HI}^{sat}$    and $\bar{x}_{HI}^{T}$   deviate by 5\% from this value. We define the deviation (in \%) as  $\Delta x_{HI}^{sat} \equiv |\bar{x}_{HI}-\overline{x}_{HI}^{sat}|/\overline{x}_{HI} = 5\%$ (column 6) and  $\Delta x_{HI}^{T} \equiv |\overline{x}_{HI}-\overline{x}_{HI}^{T}|/\overline{x}_{HI} = 5\%$ (column 7). Next, we list the values of $\bar x_{HI}^{T}$ at the point at which $dT_b/dz$ is maximal ($\bar x_{HI}^{*}$, column 8). Finally, we list $z_i$ (column 9) for which $\Delta\tau /\tau$ takes its minimal value ($\Delta\tau_{\textrm{min}}/\tau$, column 10). } \tablewidth{0pt}
\tablehead{ \colhead{Model} &  \colhead{Heating} & \colhead{$z_{\textrm{min}}$} & \colhead{$z_0$}& \colhead{$z_{\textrm{max}}$}  &  \colhead{$\overline{x}_{HI}(\Delta x_{HI}^{sat} =5\%)$} & \colhead{$\overline{x}_{HI}(\Delta x_{HI}^{T}=5\%)$}  & \colhead{$\bar x_{HI}^{*}$}& \colhead{$z_i$}& \colhead{$\Delta\tau_{\textrm{min}}/\tau$}}
\startdata 
 Massive & Hard, $f_X=0.3$ & 12.2 & 8.3 & none & 0\% & 0\%  & none & - &  $ >$1\% \\ 
& Soft, $f_X=0.3$  & 13.4 & 9.0  & 8.6 & 0\%  &  0\%  & 25\% & - & $ >$1\%\\ 
 & Hard, $f_X=1$ & 13.1 & 8.1 & none & 0\% & 0\%  & none &  - &  $ >$1\% \\ 
& Soft, $f_X=1$  & 14.2 &  10.5 & 9.2  & 0\%  &  0\%  & 31.4\% & - &  $ >$1\%\\ 
 & Hard, $f_X=30$ & 15.5 &  12.7 & 11.0  & 25.7\% &  54.9\% & 54.9\% &  $  15.3$ & 0.007\%\\ 
 & Soft, $f_X=30$  & 16.9 & 14.6 & 13.0  & 58.8\% & 60.8\%  & 64.2\% & $ 14.0$ & 0.03\%\\ 
 
Atomic & Hard, $f_X=0.3$ & 15.8 & 8.7 & 7.4 & 0\% & 0\%  & 16.3\% & - &$ >$1\% \\ 
& Soft, $f_X=0.3$  & 17.3 & 11.5 & 9.8 & 0\%  & 0\%   & 30.6\% & 16.3 & 0.01\%\\ 

 & Hard, $f_X=1$ & 16.9 & 10.9 & 9.2 & 0\% & 22.6\%  & 24.1\% & - & $ >$1\%\\ 
& Soft, $f_X=1$  & 18.3 & 13.5 &11.0 & 40.1\%  & 61.0\% & 36.7\% & 15.9 & 0.09\%\\ 
 & Hard, $f_X=30$ & 20.1 &  16.5 & 13.2 & 75.9\% &  93.5\%   & 56.6\% & $14.8$  & 0.04\%\\ 
 & Soft, $f_X=30$  & 21.4 & 18.6 & 15.0  & 87.4\% & 90.0\%  & 71.1\% & $15.3$& 0.01\%\\ 
 
  Molecular & Hard, $f_X=0.3$& 21.0 & 11.3 & 9.2 &1.8\% &2.2\% & 14.9\% & - & $ >$1\%\\ 
& Soft, $f_X=0.3$  & 22.4 & 15.0 & 12.2  &1.7\% & 1.7\% &  25.6\% & - & $ >$1\% \\ 
  & Hard, $f_X=1$& 22.4 & 14.5& 12.0 &1.8\% &26.6\% & 25.8\% & -& $ >$1\%\\ 
& Soft, $f_X=1$  & 23.9 & 17.6 & 14.5  & 41.0\%  & 46.5\%  & 34.9\% & 26.8& 0.08\%\\ 
 & Hard, $f_X=30$ & 26.6 &  21.9 & 17.5 & 78.8\% &  82.9\% & 48.8\% &  $ 24.0$  & 0.1\%\\ 
 & Soft, $f_X=30$  & 27.8 & 24.0 & 19.2 & 89.0\% & 89.0\%   & 60.0\% & $ 24.4$  & 0.03\%\\ 
\enddata
\label{Tab:1}
\end{deluxetable*}

As Figure \ref{Fig:1} suggests (and as was recently reported by \citet{Fialkov:2014N}), the saturated heating assumption may be justified only in the case of high $f_X$ (green lines in the Figure) where the IGM is indeed hotter than the CMB at the beginning of the EoR. In other  cases the gas is colder than the CMB at the beginning of reionization and undergoes the heating transition during the EoR.  The most interesting case is that of massive halos, which is also well-motivated by the low optical depth measurements. For this star formation scenario  heating is slower than reionization and the neutral gas is always colder than the CMB  in two out of six cases, namely the case of hard X-ray sources  with standard and low heating efficiency.

\section{Extracting the Neutral fraction from the global 21-cm signal} 
\label{Sec:rec}

We  would now like to mimic a global 21-cm experiment, assuming  the foregrounds fully under control, where we rely on \citet{Liu:2015} who showed that foreground contamination from Galactic synchrotron emission \citep{deOliveira-Costa:2008}, can be mitigated, allowing precise reconstruction of the optical depth from the global 21-cm signal. We first examine to which extent the global 21-cm signal can be used to constrain  the ionization history and reconstruct the neutral fraction, and then (in the next Section) use this information to extract  the total CMB optical depth.  

We start by adopting the saturated heating assumption. Given the data, $\overline {\delta T_b}$, we estimate the neutral fraction  from Eq. (\ref{Eq:2})  excluding temperature effects
\begin{equation}
\overline{x}_{HI}^{sat}\equiv \frac{\overline{\delta T_b}}{\delta T_{b,0}(1+z)^{1/2}}
\end{equation}
and check up to which values of  $\bar x_{HI}$ (listed in Table 1) the true neutral fraction is followed by the estimated one. (Following \citet{Liu:2015}, we include the factor $(1+\delta)$ into the definition of $\bar x_{HI}$, thus the quantity   $\overline{x}_{HI}$ is, in reality,   $ \overline{x_{HI}^{sat}(1+\delta)}$. However, the effect of density fluctuations on the global signal is not very large and omitting this contribution would not alter our conclusions.)

\begin{figure}
\includegraphics[width=3.4in]{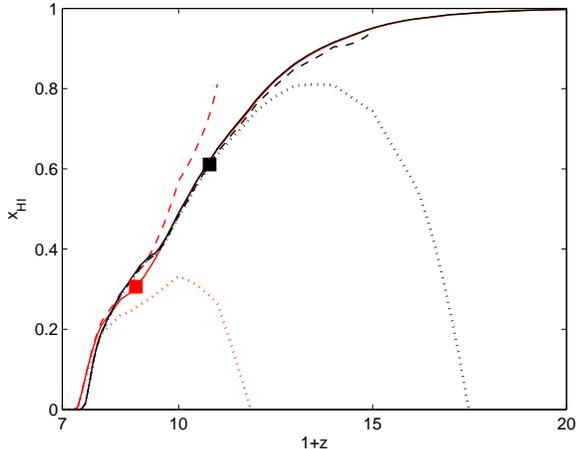}
\caption{The  reionization history for the  atomic cooling model with hard X-rays, $f_X =1$ (red) and $f_X = 30$ (black). We show the true, $\bar x_{HI}$ (solid), and the estimated, $\bar x_{HI}^{sat}$ (dotted) and $\bar x_{HI}^{T}$ (dashed), neutral fractions. The squares indicate up to which redshift we trust the reconstructed history $x_{HI}^{T}$ when fitting the ionization history  in Section \ref{Sec:OD}.}
\label{Fig:3}
\end{figure}

  As can be seen from the Table, the saturated heating assumption  is not accurate even in the case of high $f_X$, and, although the gas is hotter than CMB by the beginning of EoR, thermal effects continue to play a role. In particular,  for soft (hard) X-rays   $\overline{x}_{HI}^{sat}$ succeeds  to track the true reionization history from the end of EoR all the way up to $\bar x_{HI}\sim 89\%$ ($\bar x_{HI}\sim 79\%$) for molecular cooling, $\bar x_{HI}\sim 87\%$ ($\bar x_{HI}\sim 76\%$) in the case of atomic cooling and  $\bar x_{HI}\sim 59\%$ ($\bar x_{HI}\sim 26\%$) for massive halos. On the other hand, for low and standard heating efficiencies, $\overline{x}_{HI}^{sat}$ is a very poor approximation with a fractional error $\Delta \bar x_{HI}/\bar x_{HI}$ being greater than 5\% for all models except for  molecular and atomic cooling with soft X-rays and standard heating efficiency  in which case $\overline{x}_{HI}^{sat}$ follows the true neutral fraction up to $\bar x_{HI}\sim 40\%$. We show an example of the true ionization history and the saturated heating approximation in Figure \ref{Fig:3}. %This means that 21-cm alone is in general not a good tracer of $x_{HI}$, and in particular in the case of low and standard values of $f_X$ and hard SED. 

The situation can be alleviated  with  information on the thermal state of the IGM used. Assuming that the gas kinetic temperature, $T_K^{rec}$,  can be reconstructed from the global 21-cm spectrum, we can estimate the neutral fraction as 
\begin{equation}
\overline{x}_{HI}^{T} =  \frac{\overline{\delta T_b}}{\delta T_{b,0}(1+z)^{1/2}}\left(1-\frac{T_{CMB}}{T_K^{rec}},\right)^{-1},
\label{Eq:T}
\end{equation}
where we also adopted saturated  Ly-$\alpha$ coupling approximation. Eq. (\ref{Eq:T}) improves over the saturated heating assumption and  promises to be  a better tracer of  the true neutral fraction  than $\overline{x}_{HI}^{sat}$. 

\begin{figure*}
\includegraphics[width=3.4in]{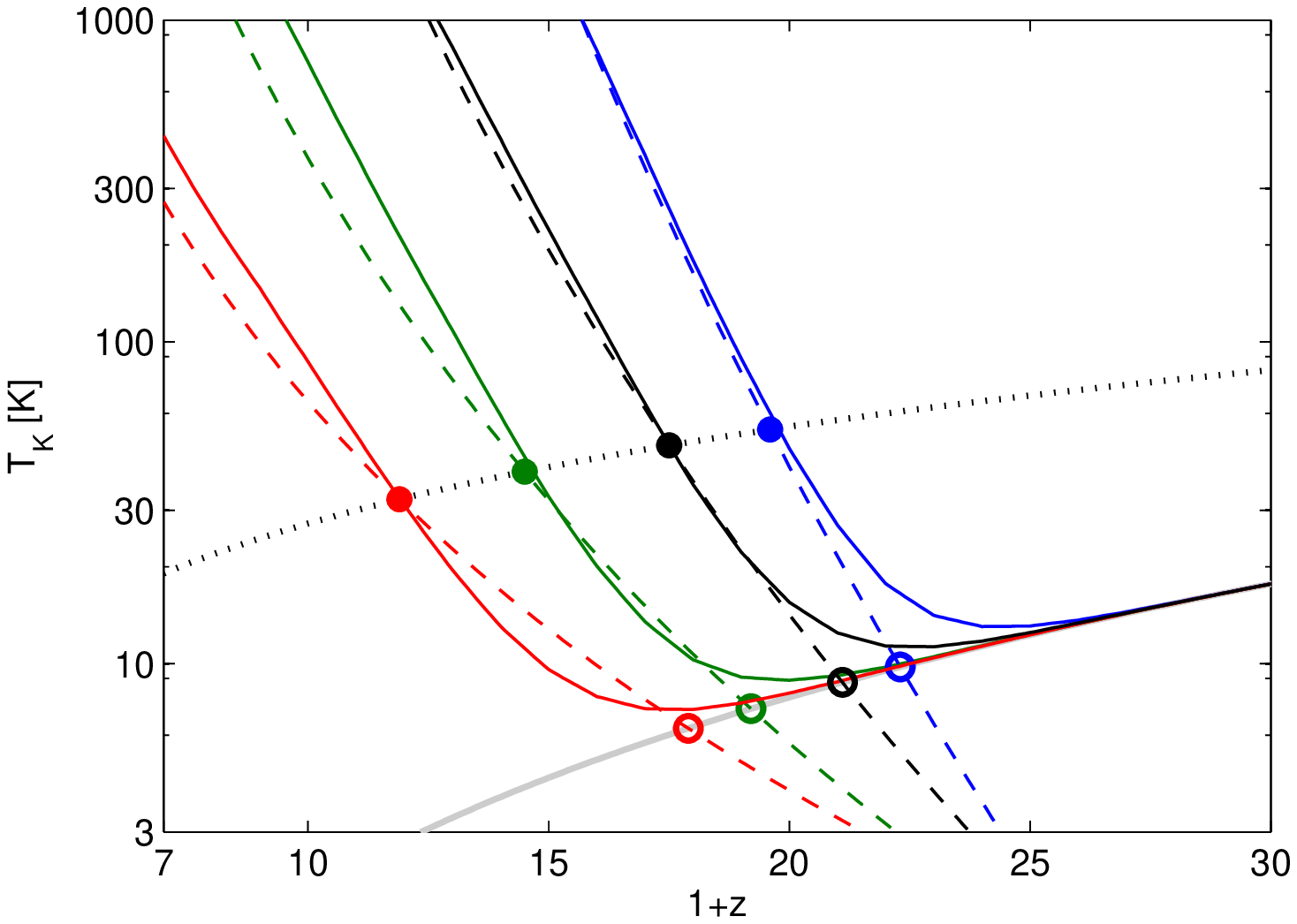}\includegraphics[width=3.4in]{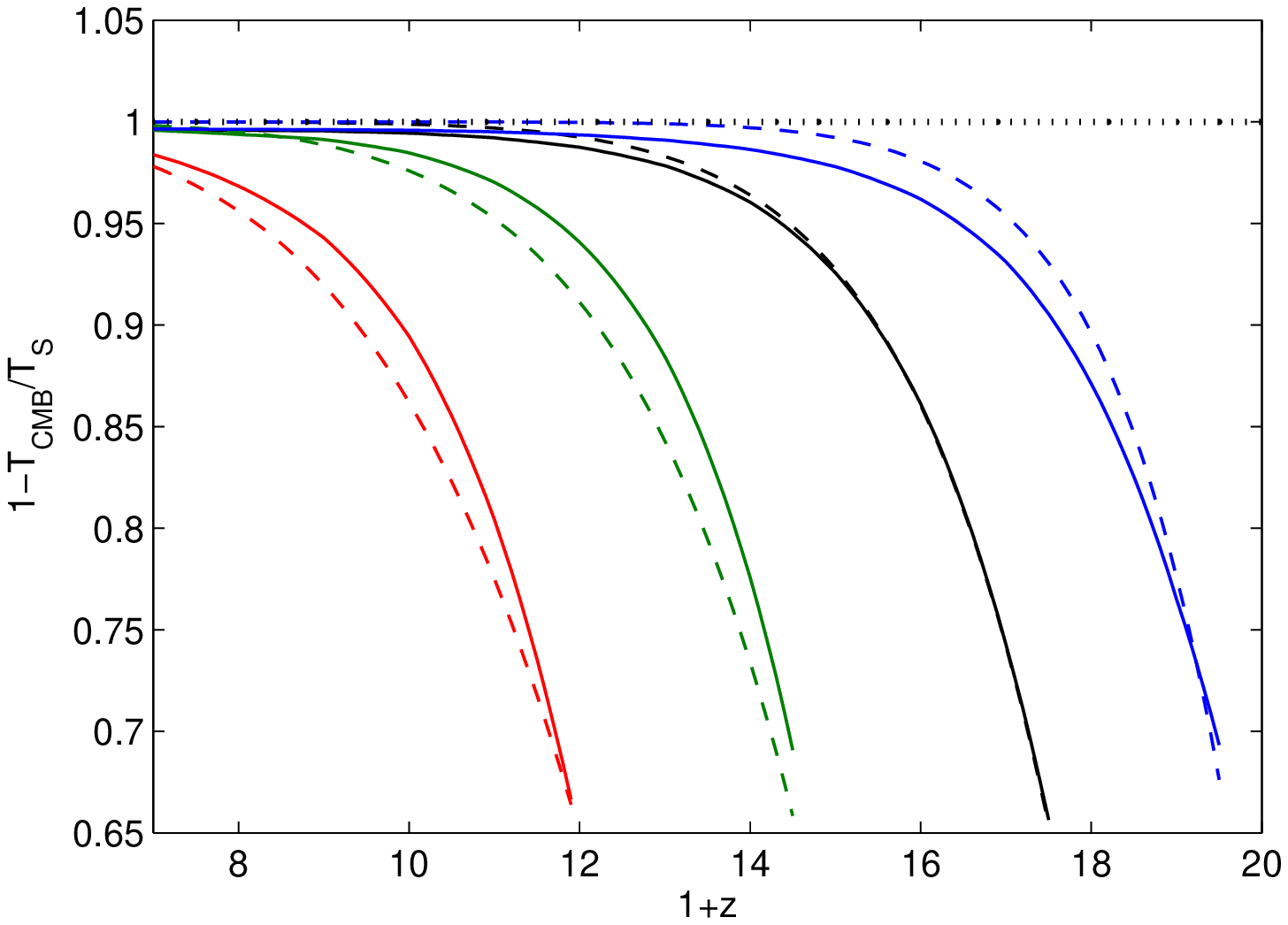}
\caption{{\bf Left}: Heating history. Solid lines show the kinetic gas temperature drawn from the simulations in the case of atomic cooling for hard SED with $f_X =1$ (red) and $f_X = 30$ (black) and soft SED for $f_X =1$ (green) and $f_X = 30$ (blue). Dashed lines show the corresponding reconstructed temperature  dependence (only the log-log interpolated piece) based in each case on two points extracted from the global 21-cm signal:  $z_{min}$ (open circles) and  $z_0$ (filled circles). The temperature of the gas which is cooled adiabatically is shown with the solid grey curve, the temperature of the CMB is shown with the dotted line. {\bf Right}: For each case from the left panel we plot  $(1-T_{CMB}/T_S)$ (solid) and $(1-T_{CMB}/T_K^{rec}) $ (dashed). In the case of saturated heating, this factor is always equal to 1 (dotted line).  }
\label{Fig:2}
\end{figure*}

As a proof of concept, we use a very simple method to extract $T_K^{rec}$ from $\overline{\delta T_b}$. Two critical points of the global spectrum  can  inform us about the heating history: (i)  the redshift of the heating transition, $z_0$, where the gas temperature equates that of the CMB, $T_{CMB} =2.725 (1+z_0)$, and  (ii) the trough of the 21-cm signal at  $z_{\textrm{min}}$  which represents the beginning of the heating era. We know that the gas cooled down adiabatically  from $z\sim 200$ to $z\sim z_{\textrm{min}}$, and, given the values of cosmological parameters, we can estimate the gas kinetic temperature at $z_{\textrm{min}}$ using publicly available codes such as RECFAST \citep{Seager:2000}. We interpolate between these two values of  redshift and temperature to reconstruct the thermal history at $z<z_{min}$ assuming  adiabatic cooling at higher redshifts. The true temperature found in our simulation and the reconstructed one are shown in Figure \ref{Fig:2} for the case of atomic cooling with $f_X = 1$ and $f_X = 30$. In the same figure we also show the  factor $(1-T_{CMB}/T_S)$ found from our mock data and compare it to the reconstructed value $(1-T_{CMB}/T_K^{rec})$ which is always equal to 1 within the saturated heating regime. Despite being a very crude approximation, $T_K^{rec}$ follows the general trend of $T_K$, and the 
reconstructed factor  $(1-T_{CMB}/T_K^{rec})$ correctly reproduces the features of the true value of $(1-T_{CMB}/T_S)$. Undoubtedly, this is a much better approximation that the saturated heating assumption; however, a better guess of the thermal history during the EoR would be very beneficial for the $x_{HI}$ extraction.
 
We use the reconstructed factor  $(1-T_{CMB}/T_K^{rec})$ to estimate the ionization history according to Eq. (\ref{Eq:T}). An example of  $\overline{x}_{HI}^T$ is shown  in Figure \ref{Fig:3}, and  we list the values of $\overline{x}_{HI}$ for which the deviation of the estimated  neutral fraction from the true one is 5\% in Table 1. With the temperature information added,  $\overline{x}_{HI}^T$ follows the true neutral fraction up to $\bar x_{HI}\sim 23\%$ in the case of  $f_X = 1$, hard SED and atomic cooling shown in the Figure (red lines), while this case was completely missed by $\bar x_{HI}^{sat}$. Moreover, for the rest of the considered models the redhsift at which the deviation reaches 5\% is pushed deeper into the first half of reionization with the exception of all the cases with $f_X = 0.3$ for which the neutral IGM is barely (or not at all)  heated to $T_{CMB}$ by the end of reionization as well as the case of massive halos with $f_X=1$ and hard SED. This inability to track $\bar x_{HI}$  is explained by the fact that our $T_K^{rec}$ is a too poor approximation and lacks  precision to serve in the regime $T_K\lesssim T_{CMB}$.

\subsection{Complete Ionization History}

Next, we would like to develop a model-independent method to reconstruct the entire reionization history based on the global 21-cm signal. To this end, we choose to use  $\overline{x}_{HI}^{T}$ (and not $\bar x_{HI}^{sat}$) as a tracer of the neutral fraction. From our analysis in the previous Section, we know that this approximation works well during the late stages of the EoR; however, we do not have a good  measure for the critical redshift (or the value of the neutral fraction), $z_*$ ($x_{HI}^*$), up to which this approximation holds. Here we adopt a rather conservative approach, outlined below, to define this instant and to reconstruct the full ionization history, $\bar x_{HI}^{rec}$.

First, we keep all the measured data points for which $\overline{x}_{HI}^{T}$ is guaranteed  to follow the true neutral fraction starting from the end of reionizaion at $z_r$ and up to $z_*$. We adopt the next model-independent criterion to find $z_*$: if the signal is seen in emission at the advanced stages of the EoR, we search for a redshift ($z_*$) between $z_r$ and the emission peak at which the derivative $dT_b/dz$ is maximal. Intuitively, this instant  marks the change in the behavior of the global signal when it transits between ionization-driven to heating-driven evolution. Clearly, this approach does not apply to the cases with no emission feature. This definition is rather conservative, and in the cases with high degree of heating we lose some information. In particular,  $\overline{ x}_{HI}^*$ is typically lower than the value of $\overline{x}_{HI}^{T}$ where it deviates from the true neutral fraction by more than 5\% (Table 1); moreover, in these cases $\bar x_{HI}^{sat}$ works as well as $\overline{x}_{HI}^{T}$ at redshifts below $z_*$. However, this  definition of $z_*$ works very well in the cases when heating is weak and extracting the reionization information from the global signal is difficult, e.g., in the cases of $f_X = 1$ with hard SED for molecular and atomic cooling. We find that in these cases $\overline{ x}_{HI}^*$ is very close to the marginal value of $\bar x_{HI}$ at which  $\overline{ x}_{HI}^T$ ceases to be a good approximation.  In other words, when using this model-independent criterion we do succeed to retain all the useful information in the ``difficult'' cases with weak heating; while we do lose some information in the ``easy'' cases with enough heating (however, as we see in the next section, this loss does not affect our main results). 

Second, we assume that EoR starts at $z_i$ with the Universe being neutral at higher redshifts. This  "anchor" point can be determined from independent experiments, e.g., using the kinetic Sunyaev-Zeldovich effect \citep{Zahn:2012}; therefore, we do not include $z_i$  in the list of our free parameters when fitting the ionization history.  

Third, in the intermediate redshift range ($z_*<z<z_i$)  $\bar x_{HI}^{rec}$  is completed using a fitting function $F(z)$. We tried several options and  found that the best results in terms of the final optical depth estimate are achieved with a three-parameter function which  appears to fit the reionization history reasonably well for all the considered cases for which our approach can be applied (i.e., all the cases which undergo the heating transition until the end of the EoR). In particular, here we choose cumulative distribution function of Gamma distribution
 \begin{displaymath}
  F(z) = \frac{1}{b^a\Gamma(a)}\int_{0}^{z-c}t^{a-1}e^{-t/b}dt,
 \end{displaymath}
 where    $a$ is the shape parameter, $b$ is the scale parameter, and $c$ marks the end of reionization. It is worth noting that in addition to the temperature effects,  photoheating feedback complicates the fitting procedure for atomic and molecular cooling. In the presence of this feedback, the low-redshift neutral fraction does no longer follow the collapsed fraction (as it does in the case of massive halos which are immune to the photoheating feedback). For    X-ray binaries with $f_X = 1$ (red curve in Figure \ref{Fig:3})  formed in atomic cooling halos the true neutral fraction follows the collapsed fraction at high values of $\bar x_{HI}$, while  changing its behavior  at $\bar x_{HI}\sim 30\%$ due to the presence of a feature (a bump) introduced by the  photoheating feedback. In this particular case, the information which we can extract from $\bar x_{HI}^T$ is dominated by the photoheating effects and does not give us any insight on the process of reionization at higher redshifts which we try to fit. 
 
In total, our reconstructed neutral fraction, which we use in the next Section to find $\tau$, is 
\begin{equation}
\overline{x}_{HI}^{rec} =\left\{
\begin{array}{l}
\overline{x}_{HI}^{T}, ~~~z<z_*\\
F(z),~~~z_*<z<z_i.\\
1,~~~ z\geq z_i
\end{array}\right.
\end{equation}

\begin{figure}
\includegraphics[width=3.4in]{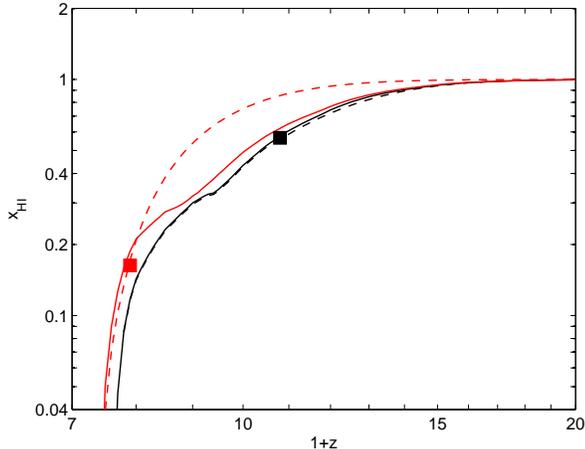}
\caption{An example of reconstructed neutral fraction (dashed) compared to $\bar x_{HI}$ (solid) for atomic cooling with $f_X = 0.3$ (red) and $f_X = 30$ (black). The squares show $z_*$ and $x_*$ for each case. Here we used $z_i = 17$ at which the true neutral fraction is $98\%$.}
\label{Fig:5}
\end{figure}

 We find that our method works well for the majority of cases with $x_{HI}^* \gtrsim 30\%$  and $F(z)$ does a decent job  reconstructing $\bar x_{HI}$ when the  starting point of reionization, $z_i$, is chosen close to the true value. Figure \ref{Fig:5} shows two examples of $\overline{x}_{HI}^{rec}$: (i) a case where  the reconstruction works well (atomic cooling with hard SED and $f_X = 30$, $x_{HI}^* \sim 57\%$, shown with black curves in the Figure), and (ii) where it fails (atomic cooling with hard SED and $f_X = 0.3$, $x_{HI}^* \sim 16\%$,  red curves). Here we clearly see that in the case of the low heating efficiency the photoheating feature is very misleading and does not allow for a more accurate fitting. 
 
  A simpler fit, such as the commonly used $\tanh(x)$ function, works well for a subset of models which we consider here, but with only two free parameters it does not capture the different shapes of the ionization history. In our case, this fit worked sufficiently well to describe the atomic cooling case with strong heating but failed to match the cases of molecular cooling and massive halos.

\section{Reconstructing the Reionization optical depth}
\label{Sec:OD}

The CMB optical depth is dependent on the ionization history 
\begin{equation}
\tau = \int (1-\bar x_{HI}) \bar n_e \sigma_Tdl,
\label{Eq:1}
\end{equation}
where $\bar n_e$ is the average number density of free electrons in ionized regions accounting for hydrogen ionization and first helium ionization, $\sigma_T$ is the Thomson cross-section and $dl$ is the line-of-sight proper distance  element.  Thus, knowing the ionization history from the global 21-cm signal should allow estimating the optical depth.% As was mentioned earlier, a possibility of reconstructing the optical depth based on global 21-cm measurements was recently discussed by \citet{Liu:2015}; however, the authors assumed that $T_S>>T_{CMB}$ implying that the IGM is much warmer than the CMB during all the process of reionization. Here we relax this assumption. 

 Although the reconstruction $\bar x_{HI}^{rec}$ does not work perfectly well to reproduce $\bar x_{HI}$ as can be seen from Figure  \ref{Fig:5}, the error in $\tau$ is expected to be much smaller than the error in $\bar x_{HI}$ itself  because: (i) the largest part of the optical depth is contributed by redshifts $z<z_r$ when the Universe was fully ionized (in our case of massive halos with reionization ending at  $z_r\sim 8$  only 30\% of the optical depth is sourced by the ionized patches during the EoR); and (ii)  the fit over- and under-predicts $\bar x_{HI}$ at different redshifts which results is partial cancellation of the error. 
 
\begin{figure*}
\includegraphics[width=2.4in]{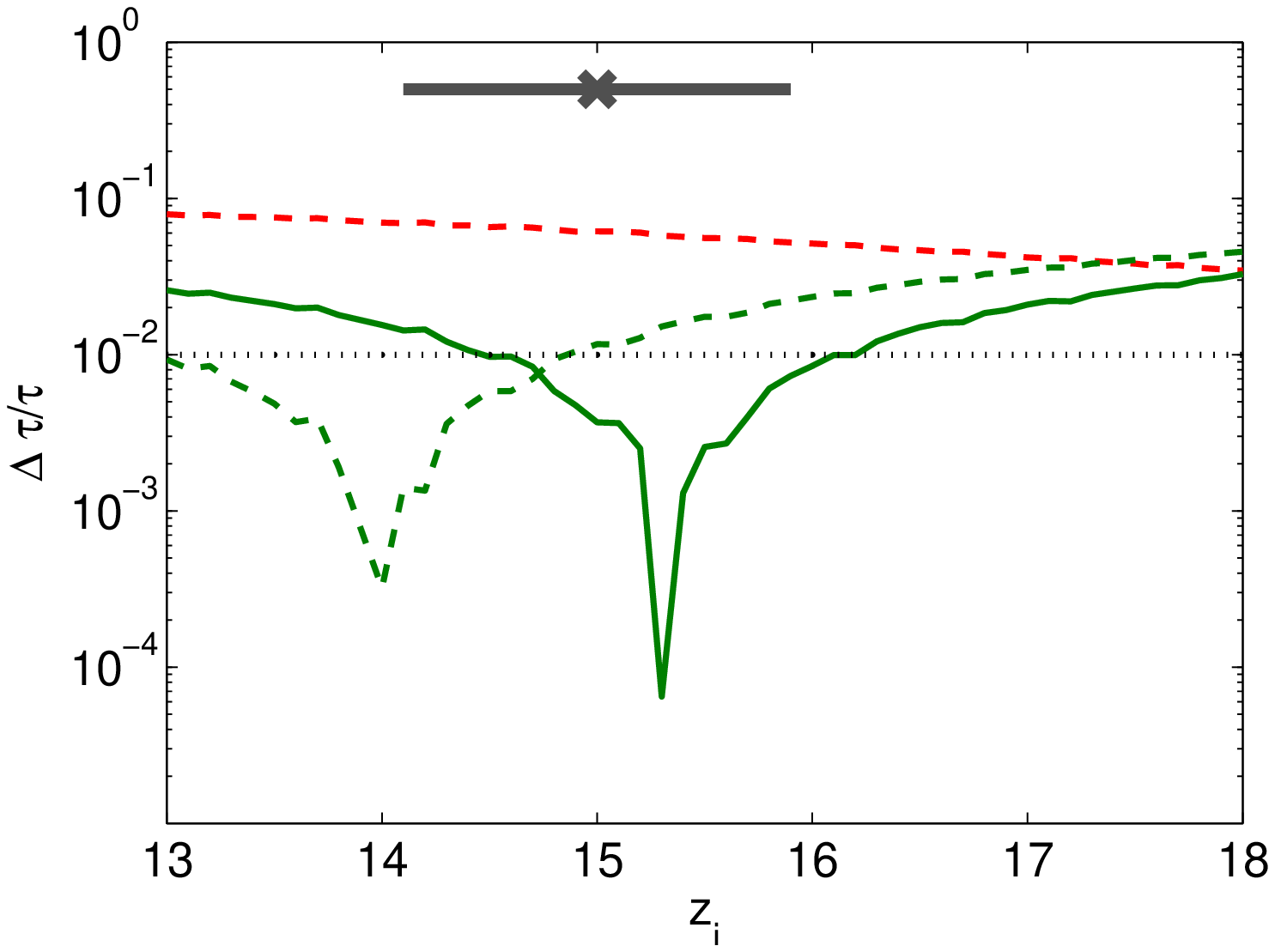}\includegraphics[width=2.4in]{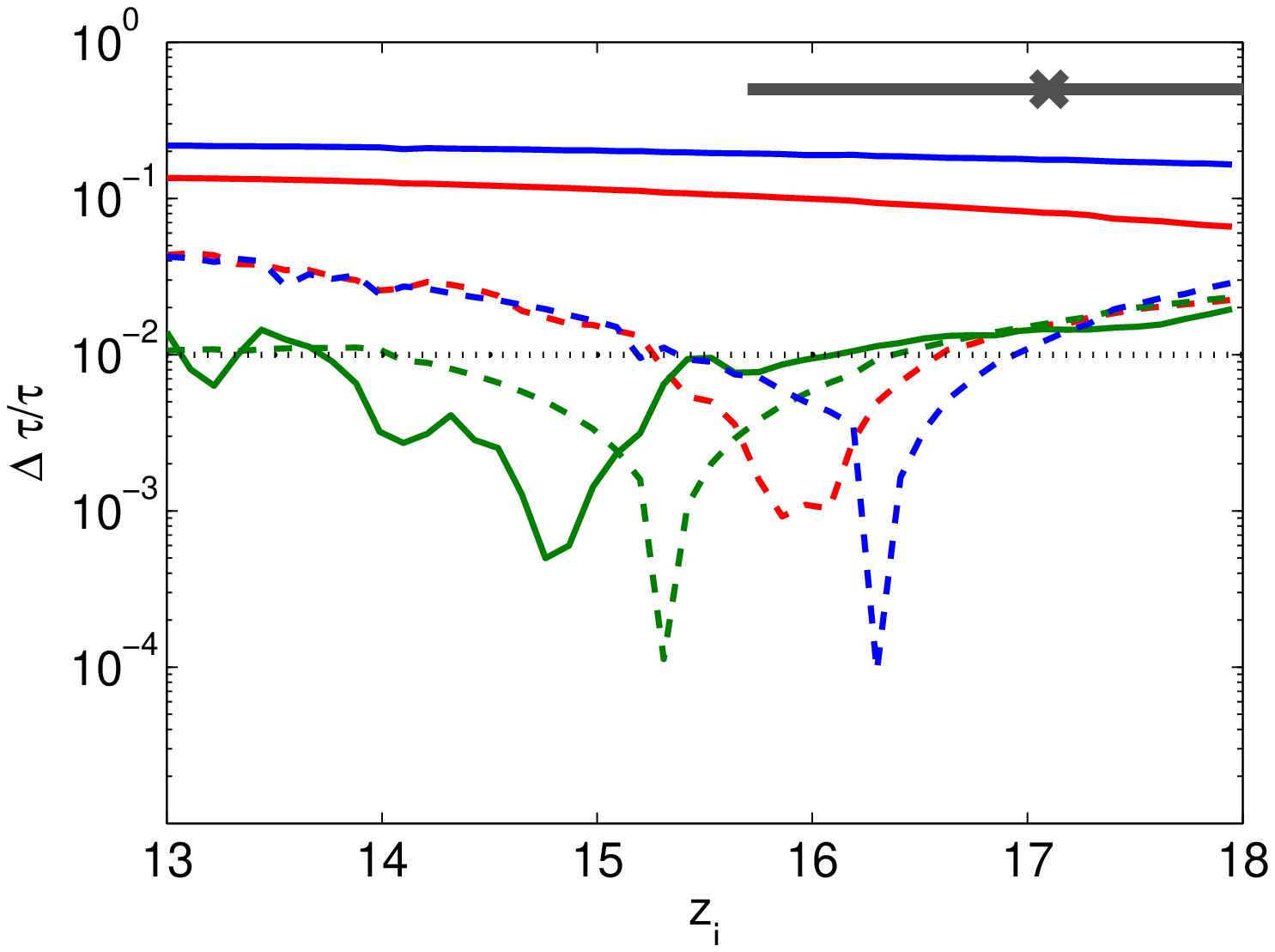}\includegraphics[width=2.4in]{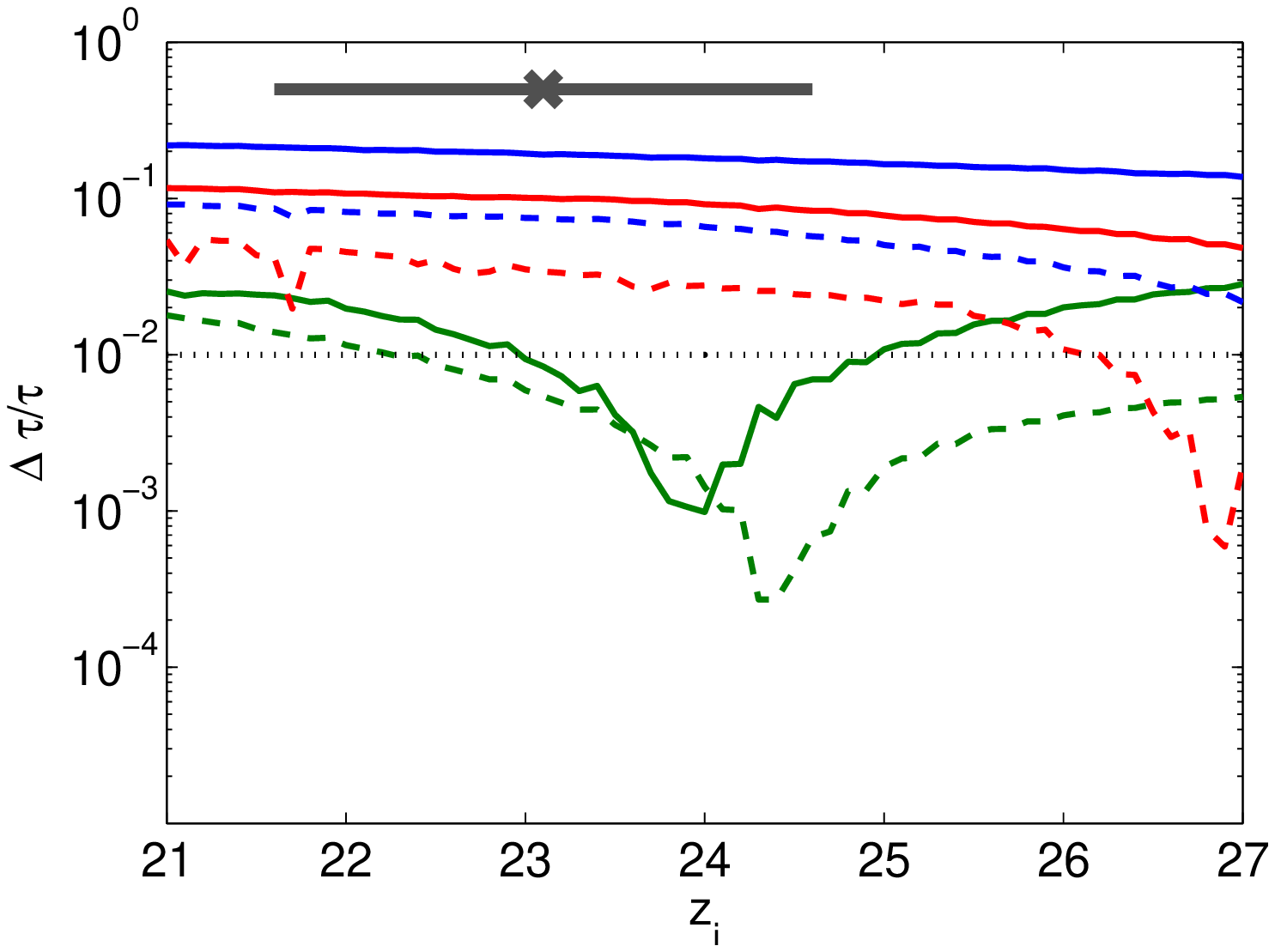}
\caption{Total relative error  $\Delta \tau/\tau$ is shown as a function of $z_i$ for   massive halos (left), atomic cooling (middle) and molecular cooling (right) scenarios with hard (solid) and soft (dashed) X-ray sources of heating efficiency $f_X = 0.3$ (blue), $f_X = 1$ (red) and $f_X = 30$ (green). The horizontal black dotted line marks the $\Delta \tau/\tau = 0.01$ threshold. We also show the true beginning of reionization in our models (shown for  hard SED with $f_X = 1$ in each case): the  thick grey bar marks 0.5\%-2\% range in ionized fraction and the cross marks 1\% ionization. Here we use a resolution of $\Delta z_i = 0.1$, i.e., our error curves are smoothed on this scales. }
\label{Fig:4}
\end{figure*}

Using $\bar x_{HI}^{rec}$ we compute the optical depth  $\tau^{rec}$ and compare it to the true value, $\tau$, found directly from the simulation data. The accuracy with which the optical depth can be extracted from the global signal depends on the value of $z_i$, as can be seen from Figure \ref{Fig:4} where  the fractional error in the optical depth, $\Delta\tau/\tau = |\tau^{rec}-\tau|/\tau$, is shown   as a function of $z_i$  for all the cases where the fitting procedure converged.  In most of  our cases $\Delta\tau/\tau$ features a broad minimum (of $\Delta z_i \sim 2$) within which  the fractional error in $\tau$ is below 1\%. The location of this feature is very close to  the true beginning of EoR, marked by grey bars in Figure \ref{Fig:4} which correspond to the  $ 0.5-2\%$  values of ionized fraction.  The minimal value of the fractional error, which we quote  in Table 1 together with the corresponding $z_i$ is below $0.1\%$, which is much better than the current 1$\sigma$ confidence level of the Planck satellite ($\sim 24\%$).  In cases where the reconstruction does not work well and  the fractional error does not feature a minimum,  $\Delta \tau/\tau$ remains below $\sim 10-20\%$ level in the  $ 0.5-2\%$  range of the ionized fraction.

\section{Conclusions}
\label{Sec:conc}

The total CMB optical depth is a long-standing nuisance for CMB cosmology. Here we have examined to which extent the global 21-cm signal can be used to probe the total CMB optical depth in realistic cases of IGM heating, including hard and soft X-ray sources with low, standard and high heating efficiency. Following \citet{Fialkov:2014N}, we have shown that the intensity of the 21-cm signal produced during the EoR is strongly affected by the thermal state of the IGM in addition to its ionization, which  makes it harder to extract the reionization history from the global 21-cm signal compared to a scenario in which heating is saturated \citep{Liu:2015}. 

We have developed a simple and model independent approach to reconstruct the neutral fraction from a realistic global 21-cm signal and used it to estimate the optical depth for a large variety of models with different ionization and heating histories. The method can be summarized as follows: (i) at low redshifts we extract the neutral fraction from the global 21-cm signal going beyond the saturated heating assumption and using information on the thermal state of the IGM  extracted directly from the mock global 21-cm signal; (ii) we assume that the redshift at which reionization starts, $z_i$, is known with the Universe neutral at that epoch; (iii) we  complement the neutral fraction in the intermediate redshift range using a three-parameter fitting function which works well for the different types of reionization histories which we have explored. 

One of the main conclusions we reach is that with the thermal history added a better estimation of the reionization history is possible, and  the neutral fraction can be reconstructed even when the 21-cm signal is affected by thermal history all the way throughout the  EoR. As a proof of concept,  we adopt a very simple method to estimate the temperature of neutral IGM using two critical points of the global signal, namely (i) the heating transition at which the gas kinetic temperature equates that of the CMB, and (ii) the beginning of the heating era when X-ray sources turn on. Even this simple method improves over the saturated heating approximation. 

Finally, we calculate the optical depth using the extracted reionization history and show that an accurate  measurement of $\tau$, with fractional error below 1\% over a wide range of $z_i$, is possible even when the IGM heating is not saturated all the way throughout the EoR. We have blindly tested our method on a large variety of ionization histories for different star formation scenarios varying the low-mass cutoff of star-forming halos.

Our results  are  timely considering the plethora of existing and planned global 21-cm experiments which might remove  the optical depth nuisance from the CMB cosmology in near future, allowing for a much more precise determination of the cosmological parameters.

\acknowledgments

We thank R. Barkana and A. Cohen for their contribution to preceding works which provided a solid basis for this paper. We thank R. Barkana for his valuable comments on the draft of this paper. This work was supported in part NSF grant AST-1312034 (for A.L.). 


%% in parentheses, after the keyword, but they will not be verified.



\end{document}